# Time delay signature elimination of chaos in a semiconductor laser by dispersive feedback from a chirped FBG

Daming Wang, Longsheng Wang, Tong Zhao, Hua Gao, Yuncai Wang, Xianfeng Chen, and Anbang Wang*, *Member, IEEE*

*Abstract*—Time delay signature (TDS) of a semiconductor laser subject to dispersive optical feedback from a chirped fiber Bragg grating (CFBG) is investigated experimentally and numerically. Different from mirror, CFBG provides additional frequency-dependent delay caused by dispersion, and thus induces external-cavity modes with irregular mode separation rather than a fixed separation induced by mirror feedback. Compared with mirror feedback, the CFBG feedback can greatly depress and even eliminate the TDS, although it leads to a similar quasi-period route to chaos with increases of feedback. In experiments, by using a CFBG with dispersion of 2000ps/nm, the TDS is decreased by 90% to about 0.04 compared with mirror feedback. Furthermore, both numerical and experimental results show that the TDS evolution is quite different: the TDS decreases more quickly down to a lower plateau (even background noise level of autocorrelation function) and never rises again. This evolution tendency is also different from that of FBG feedback, of which the TDS first decreases to a minimal value and then increases again as feedback strength increases. In addition, the CFBG feedback has no filtering effects and does not require amplification for feedback light.

## I. INTRODUCTION

Chaotic semiconductor lasers with external-cavity feedback have attracted widespread attention due to complex dynamics, simple structure, and promising applications, including secure communication [1, 2], random number generation [3-7], radar [8-10], and chaos optical time-domain reflectometry [11, 12]. Complexity is one of the attractive features of the laser chaos. Unfortunately, it has been proven that the laser with external mirror feedback has the signature of time delay, i.e., the round-trip time between the laser and the mirror [13, 14]. One of the typical characteristics of the laser chaos is that the autocorrelation function (ACF) of chaotic waveform has a peak located at the feedback delay, of which the height is usually used to quantitatively estimate as the time delay signature (TDS). This means that the laser chaos has periodicity and thus the complexity is limited. Moreover, the TDS divulges the information of length of the external cavity and increases the possibility of system reconstruction [15]. Resultantly, the performances of applications of chaotic external-cavity laser are seriously restricted by the TDS, such as the security of chaos communications, the randomness of random number generation, the ambiguity and anti-jamming of radar finding.

Rontani *et al* first pointed out the TDS of external-cavity laser, and proposed suppressing TDS by adjusting the feedback strength and the bias current such that the laser relaxation-oscillation period is close to the time delay [13, 14]. Due to that the laser relaxation period is changed in a small range by adjusting current or feedback strength, this method is only suitable for short feedback cavity comparable to the relaxation period. Thereafter, other different approaches were proposed [16-33]. These approaches can be classified into two types. One is increase of the configuration complexity of feedback cavity, including double cavity feedback with two mirrors [16], combing feedback with optical injection from an additional laser [17, 18], modulation of multiple delays [19], and polarization-rotated or polarization-resolved feedback (only for VCSEL) [20-22]. The other type depresses the TDS by replacing the external mirror by a nonlinear device, such as an additional laser [23, 24] and an optical filter [25-28]. The TDS suppression relies on that the nonlinear response of feedback device reduces the correlation between the

This work was supported in part by National Nature Science Foundation of China (NSFC) under grants 61475111, 61671316, in part by Natural Science Foundation for Excellent Young Scientists of Shanxi (2015021004); in part by Program for the Innovative Talents of Higher Learning Institutions of Shanxi; and in part by International Science and Technology Cooperation Program of China and Shanxi (2014DFA50870, 201603D421008).

The authors except XF Chen are with the Key Laboratory of Advanced Transducers and Intelligent Control System, Ministry of Education and Shanxi Province, and are also with College of Physics and Optoelectronics, Taiyuan University of Technology, Taiyuan 030024, China; XF Chen is with the School of Electronic Engineering, Bangor University, Bangor LL571UT, UK (*Corresponding author: Anbang Wang, e-mail: wanganbang@tyut.edu.cn)

feedback light and the laser output. For the laser as feedback device, the nonlinear response comes from the dynamics induced by unlocking-injection. Laser parameters need to be optimized such that the mutually coupled lasers operate at a small unlocking regime [23]. For filtered optical feedback, Li *et al* numerically and experimentally achieved TDS suppression by using a fiber Bragg grating as feedback device [26, 27]. The nonlinear response originates from the dispersion-induced group delay. Due to that the dispersion is close to zero in the main lobe of FBG reflectivity spectrum, the laser spectrum needs to be tuned to edges of the main lobe to find a minimal TDS by optimizing the laser frequency detuning and the feedback strength [27]. In addition, this sideband filtering reflection leads to power loss and thus an optical amplifier is required in the feedback cavity to ensure sufficient feedback strength [28], which makes the configuration complex.

In this paper, we numerically and experimentally demonstrate a semiconductor laser with dispersive optical feedback from a chirped FBG (CFBG) for eliminating the TDS. In this method, the dispersion and the induced group delay occur in the main lobe of CFBG reflectivity spectrum. It should be mentioned that, the frequency-detuned FBG feedback actually induces frequency-related reflectivity (or feedback strength) and thus belongs to optical filtered feedback. By contrast, in the main lobe of CFBG spectrum, the reflectivity is independent of frequency, and therefore the CFBG feedback has no filtering effects and is a typical paradigm of dispersive feedback. In CFBG feedback, it is no longer required to optimize parameters to tune the laser center spectrum to the sideband of grating. In addition, the filtering-induced optical loss can be greatly decreased which is helpful to simplify system configuration. The numerical and experimental results show that the TDS decreases more quickly down to a lower plateau, even to the ACF background noise for sufficient dispersion, and never rises again, which is different from mirror feedback and FBG feedback. In addition, our numerical results reveal that the elimination of TDS is related to the irregular external cavity modes caused by dispersive feedback. In experiments, using a CFBG with dispersion of about 2000ps/nm, a minimal TDS of 0.04 is achieved, which is decreased by 90% compared to mirror feedback. According to numerical results, the TDS can be perfectly removed by further increases of dispersion.

## II. NUMERICAL MODEL AND RESULTS

*A. Model of semiconductor laser with dispersive feedback*

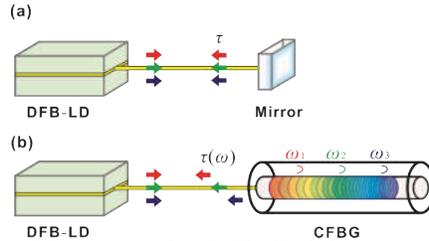

Fig. 1. The schematic diagrams of a semiconductor laser subject to external optical feedback from (a) a mirror and (b) a chirped FBG (CFBG).

Figure 1 shows the schematic configurations of the mirror-feedback and the CFBG-feedback laser systems. The only change in configuration is replacing the mirror by the CFBG. Then, the semiconductor laser subject to dispersive feedback from a CFBG is modeled by the following rate equations, which are obtained by modifying the Lang-Kobayashi model [29].

$$\frac{dE}{dt} = \frac{1+i\alpha}{2}[\frac{g(N-N_0)}{1+\varepsilon|E|^2} - \frac{1}{\tau_p}]E + \frac{\kappa_f}{\tau_{in}}\int_{t-T}^{t} h(t-t')E(t'-\tau)dt' \quad (1)$$

$$\frac{dN}{dt} = \frac{I}{qV} - \frac{N}{\tau_N} - \frac{g(N-N_0)}{1+\varepsilon|E|^2}|E|^2 \quad (2)$$

Where $E$ is the complex amplitude of electrical field of the laser and $N$ is the carrier density in laser cavity. The second term in the right side of Eq. (1) represents the dispersive optical feedback. The integral in the feedback term is the convolution between the response function $h(t)$ of the CFBG and the delayed laser field $E(t-\tau)$. Parameter $\tau$ is the round-trip time between laser and grating, $\kappa_f$ is the amplitude feedback strength, and $T$ is the integral length which should be larger than the response time of grating. Note that for mirror feedback, $h(t)$ is an impulse function and then the integral equals $E(t-\tau)$. For a CFBG, the response function $h(t)$ does not have an analytic expression. Therefore, based on convolution theorem, the dispersive feedback term is numerically calculated by inverse Fourier transform of $H(\omega)\cdot\mathbf{FT}\{E(t-\tau)\}$, where $\mathbf{FT}\{\}$ denotes Fourier transform and $H(\omega)$ is the complex reflection spectrum of the chirped grating.

The chirped grating spectrum is obtained by the piecewise-uniform approach [30], which is outlined as follows. The chirped grating is equally divided into $M$ small pieces and each piece is considered as a uniform grating. Based on the propagation matrix method, the amplitudes of the forward-going and backward-going waves $S_M$ and $R_M$ at the left interface (i.e., incidence interface) of the chirped grating can be obtained as

$$\begin{bmatrix} S_M \\ R_M \end{bmatrix} = \mathbf{F}_M \cdot \mathbf{F}_{M-1} \cdot \ldots \cdot \mathbf{F}_j \cdot \ldots \cdot \mathbf{F}_1 \begin{bmatrix} 1 \\ 0 \end{bmatrix}. \quad (3)$$

The reflection spectrum is then calculated as $R_M/S_M$. In Eq. (3), $\mathbf{F}_j$ is the propagation matrix of the $j^{th}$ sub-grating and it is expressed as follows according to the coupled-mode theory,

$$\mathbf{F}_j = \begin{bmatrix} \cosh(\gamma_j \Delta L) - i\dfrac{\hat{\sigma}_j}{\gamma_j}\sinh(\gamma_j \Delta L) & -j\dfrac{\kappa_j}{\gamma_j}\sinh(\gamma_j \Delta L) \\ i\dfrac{\kappa_j}{\gamma_j}\sinh(\gamma_j \Delta L) & \cosh(\gamma_j \Delta L) + i\dfrac{\hat{\sigma}_j}{\gamma_j}\sinh(\gamma_j \Delta L) \end{bmatrix} \quad (4)$$

with

$$\gamma_j = \sqrt{\kappa_j^2 - \hat{\sigma}_j^2},$$
$$\hat{\sigma}_j \approx n_{\text{eff}}(\omega - \omega_j)/c, \quad \kappa_j = \omega_j \rho <\delta n>/2c,$$

where $\omega_j = \pi c/n_{\text{eff}}\Lambda_j$, $\Delta L$ and $\Lambda_j$ are the Bragg frequency, grating length and grating period of the $j^{th}$ uniform grating, respectively, $n_{\text{eff}}$ is the effective index, $<\delta n>$ is the average index change, $\rho$ is the fringe visibility of the index change in grating, and $c$ is the speed of light in vacuum.

In our simulation, the chirped grating parameters are $n_{\text{eff}}=1.46$, $<\delta n>=5\times10^{-4}$, $\rho=1$, grating length $L=10$cm, and grating center wavelength 1550nm. To model reflection spectrum and delay spectrum, the chirped grating is divided into 200 sub-gratings and the Bragg frequencies $\omega_j$ of these sub-gratings are calculated according to chirp factor $d\lambda/dz$ (nm/cm). Note that the grating dispersion can be read from the delay spectrum. Back to Eqs. (1) and (2), the laser parameters are listed as follows: transparency carrier density $N_0=1\times10^6$ μm$^{-3}$, differential gain $g=0.5625\times10^{-3}$ μm$^3$ns$^{-1}$, gain saturation parameter $\varepsilon=1\times10^{-5}$ μm$^3$, carrier lifetime $\tau_N = 2.2$ns, photon lifetime $\tau_p = 1.17$ps, linewidth enhancement factor $\alpha=6.0$, round-trip time in laser cavity $\tau_{in} = 7.3$ps, round-trip time between the laser and grating $\tau = 5$ns, active layer volume $V=100$μm$^3$, and the elementary charge $q=1.602\times10^{-19}$C, and the laser bias current $I=1.5I_{th}$, where $I_{th}=15.35$ mA is laser threshold current. The integral time $T$ for calculating dispersive feedback is set as 20.48ns. In the following text, the TDS is depicted by the ACF and the delayed mutual information (DMI) traces of laser intensity waveform [25] and is quantitatively evaluated as the height of ACF peak in a small interval around the time delay.

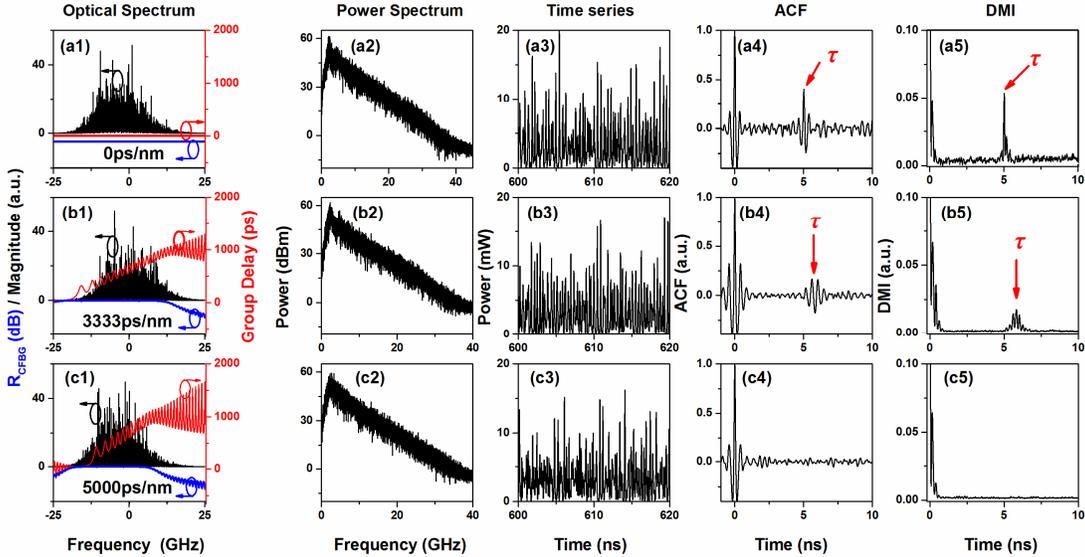

Fig. 2. Numerically simulated optical spectra (column 1), power spectra (column 2), time series (column 3), ACF traces (column 4) and DMI traces (column 5) with different dispersion: (a1-a5) 0ps/nm, i.e., mirror feedback, (b1-b5) 3333ps/nm and (c1-c5)5000ps/nm. The delay spectra and reflection spectra of CFBG are plotted with optical spectra together. $I = 1.5I_{th}$, $\kappa_f=0.1$, $\tau=5$ns.

*B. Numerical results*

Figure 2 shows the simulated time- and frequency-domain characteristics of semiconductor laser subject to CFBG feedback with different dispersion values of 0, 3333, and 5000ps/nm under a fixed feedback strength $\kappa_f=0.1$. From left to right, columns 1-5 list optical spectra, power spectra, time series, ACF and DMI of the time series, respectively. In the first column the reflection spectra and delay spectra of chirped grating are also plotted in dark gray (blue) and gray (red), respectively. The first row shows the results of zero-dispersion feedback, i.e. mirror feedback. As shown in Fig. 2(a1), all optical spectral components have no additional time delay due to mirror feedback. The laser has a typical chaotic intensity output which has a sharp power spectrum dominated by relaxation oscillation and a large-amplitude irregular temporal waveform plotted in Figs. 2(a2) and 2(a3), respectively. Then, the TDS can be clearly found in ACF trace as well as in DMI trace. As shown in Fig. 2(a4) and Fig. 2(a5), the peak values at the time delay are 0.47 and 0.05, respectively. As the dispersion increases to 3333ps/nm, as shown in Fig. 2(b1), the optical spectral components have different additional delay caused by the chirped grating. In this case, although the profile of power spectrum (Fig. 2(b2)) and the time series (Fig. 2(b3)) are similar to those induced by mirror feedback, the TDS is suppressed, shown by ACF and DMI traces in Figs. 2(b4) and 2(b5). Quantitatively, the size of the ACF peak and DMI peak are decreased by 65%. It should be noticed that the position of the ACF peak shifts right about 0.5ns, which is caused by the additional delay of the strongest mode. The bottom row shows the results with dispersion of 5000ps/nm. It is found that there is no peak in the ACF and DMI traces from Figs. 2(c4) and 2(c5). This demonstrates that the dispersion feedback from CFBG can eliminate the TDS.

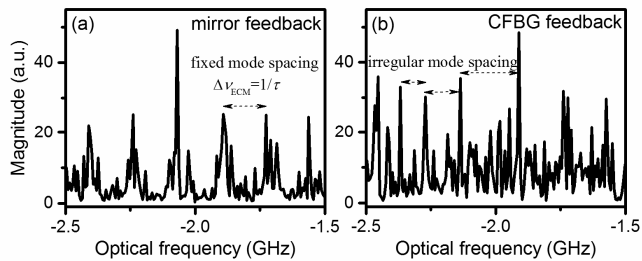

Fig. 3. Simulated local optical spectra of laser subject to (a) mirror feedback and (b) CFBG feedback with dispersion of 5000ps/nm, corresponding to Fig. 2(a1) and Fig. 2(c1), respectively. $I = 1.5I_{th}$, $\kappa_f=0.1$, $\tau =5$ns.

To understand effects of dispersive feedback on TDS elimination, we analyze the change of external-cavity modes (ECMs). Figures 3(a) and 3(b) plot the magnified local optical spectra of laser with mirror feedback and CFBG feedback, corresponding to Figs. 2(a1) and 2(c1), respectively. For mirror feedback, the resonant ECMs $v_m$ satisfy resonance condition $v_m\tau=m$, where $m$ is the order number and $\Delta m=1$ for neighboring modes, and then the mode separation is equal to $1/\tau$ as shown in Fig. 3(a). These modes have the least common multiple period, i.e., the same round-trip period and thus lead to the TDS [31]. By contrast, for CFBG feedback, the dispersion results in frequency-related additional delay time. Resultantly, the solution of resonance condition becomes more complex. Qualitatively, it can be deduced that the mode spacing is no longer a fixed value, which is numerically shown in Fig. 3(b). Therefore, the ECMs caused by CFBG do not have the least common multiple period and then the TDS is eliminated.

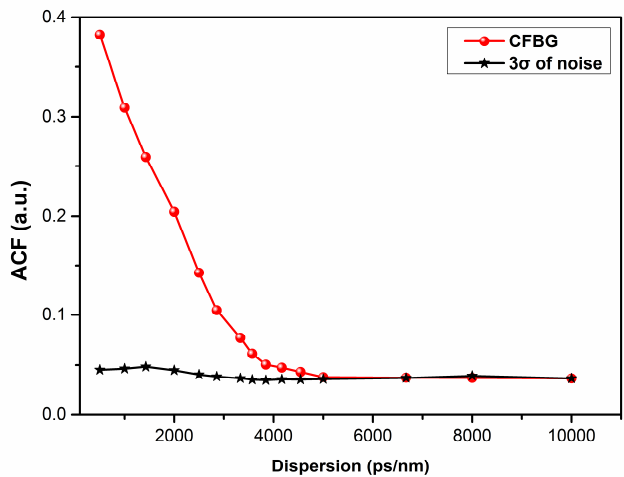

Fig. 4. Numerically obtained effect of CFBG dispersion on TDS. $I = 1.5I_{th}$, $\kappa_f=0.1$, $\tau =5$ns, $L=10$cm. $\sigma$ is the standard deviation of the background noise of ACF traces.

Figure 4 shows the effect of CFBG dispersion on the TDS which is numerically obtained at $\kappa_f= 0.1$, i.e., 1% of intensity feedback level. The stars represent the three standard deviation of the background noise of ACF traces. It is found that the TDS is gradually reduced as increasing the dispersion value of CFBG, and eventually decreases down to background noise as dispersion exceeds about 5000ps/nm. There is a critical dispersion value for eliminating the TDS; however it is difficult to obtain the quantitative expression of the critical dispersion. Note that the maximum delay difference among oscillated modes can be evaluated as the product of the grating dispersion and the optical spectral width. If the maximum delay difference is smaller than the correlation time of laser output, all the modes are somewhat correlated, and thus the TDS cannot be depressed effectively. Therefore, it is extrapolated that the critical dispersion should be proportional to the correlation time of laser output and also be inversely proportional to the optical spectral width.

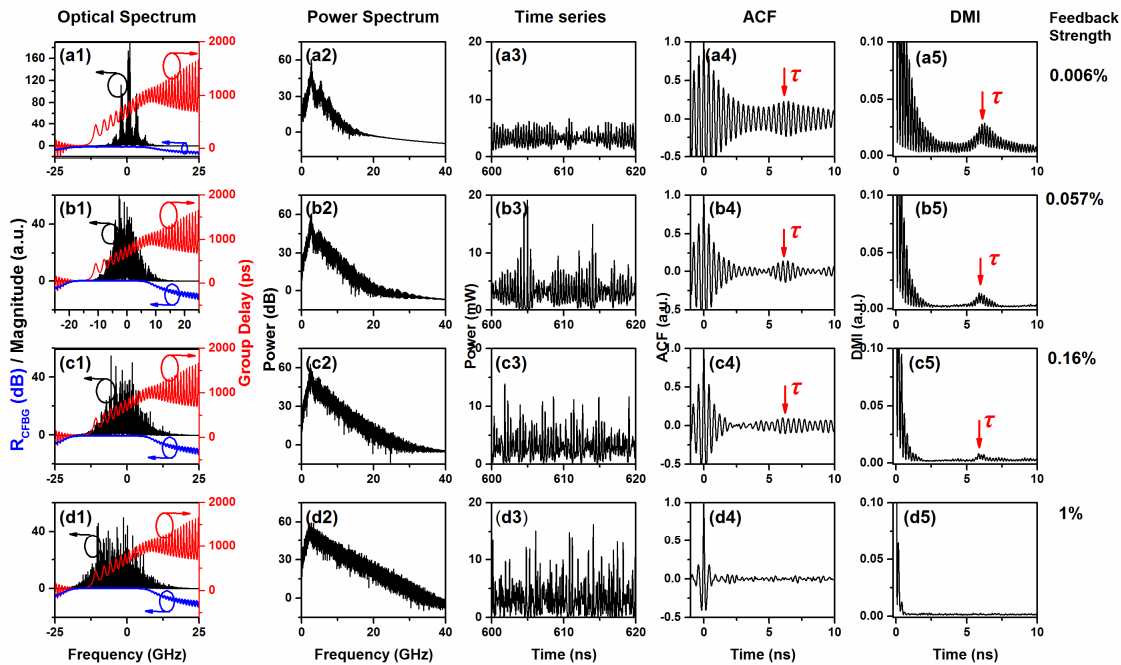

Fig. 5. Numerically simulated output of laser with 5000ps/nm CFBG feedback separately obtained at intensity feedback levels of 0.0065% (a1-a5), 0.057% (b1-b5), 0.16% (c1-c5), and 1% (d1-d5). In each row, optical spectrum, power spectrum, time series, ACF trace and DMI trace are plotted from left to right. The reflection and delay spectrum of CFBG are plotted in the first column.

It is well known that feedback strength affects the optical spectral width. Therefore we further study the effects of feedback strength on the TDS of dispersive CFBG feedback. Figure 5 demonstrates the output characteristics of the laser separately simulated with intensity feedback levels (i.e., $\kappa_f^2$) of 0.006%, 0.057%, 0.16%, and 1%. In this figure, columns 1-5 list optical spectra, power spectra, time series, ACF and DMI of the time series, respectively, from left to right. For a weak feedback of 0.006%, only relaxation oscillation is excited as shown in Fig. 5(a1), and then the laser exhibits a quasi-period oscillation at the laser relaxation frequency, which is shown by the power spectrum in Fig. 5(a2). The ACF trace in Fig. 5(a4) as well as the DMI trace in Fig. 5(a5) has a local maximum at about 6 ns which reveals the TDS. The ACF peak and DMI peak are 0.23 and 0.03, respectively. The shift of TDS from 5ns is caused by the dispersion-induced additional delay to the center modes. It is worth noting that the ACF in Fig. 5(a4) decreases with a damped periodic oscillation but it does not decrease to zero before 5ns, i.e., the time delay. This indicates that the correlation induced by the laser's quasi-period oscillation also contributes to the TDS. As feedback level increases, for instance, to 0.057% or 0.16%, more modes are excited by feedback as depicted in Fig. 5(b1) and Fig. 5(c1). Thus, the laser dynamics becomes more complex, as depicted by the corresponding power spectra and time series. As plotted in Figs. 5(b4) and 5(c4), the ACF's periodic oscillation is quickly damped to zero, and thus the TDS is decreased. Further increase of feedback level leads to a developed chaotic oscillation, of which the optical spectrum (Fig. 5(d1)) is broadened greatly. In addition, the relaxation oscillation is depressed in laser dynamics; as shown in Fig. 5(d4), the ACF is damped to zero only after one relaxation period. Interestingly, in this case there is no peak appearing between 5~7ns. This means that the TDS is eliminated caused by dispersive feedback. By comparing the optical spectra of the four feedback levels, we can find that the lager the spectral width is, the much greater the TDS suppression is.

Figure 6 shows the numerically calculated TDS as function of feedback level at different dispersion values of 2500ps/nm (squares), 3333ps/nm (dots), 5000ps/nm (triangles) and 10000ps/nm (diamonds). The results of the four dispersion values show a similar evolution tendency. As feedback level increases, the TDS is first reduced quickly to a knee point and then slowly down to the background noise if the dispersion is sufficient. This tendency is different from that of mirror-feedback laser system, which will be shown in the following experimental results. Furthermore, the TDS tendency is also different from that of the uniform FBG feedback which has been shown in [27]. For uniform grating feedback, the laser spectrum needs to be detuned to edges of the main lobe of reflection spectrum. As a result, the frequency-detuned FBG response has the filtering effects and thus is very sensitive to the shifting or broadening of the laser optical spectrum. It is well known that increasing feedback strength leads to spectrum broadening. Therefore, after reaching a minimal value, the TDS of FBG feedback rises again as further increasing feedback strength [27]. By comparison, as shown in Fig. 2(c1) and Fig. 5(d1), the TDS elimination in CFBG feedback is readily achieved when the laser optical spectrum is within the main lobe of the reflection spectrum of the chirped grating. Thus, the CFBG response does not contain the filtering effects. During feedback strength increases, the laser center spectrum is always within the reflection spectrum of CFBG, and resultantly the TDS suppression is kept. This is the distinguishing feature of the CFBG feedback.

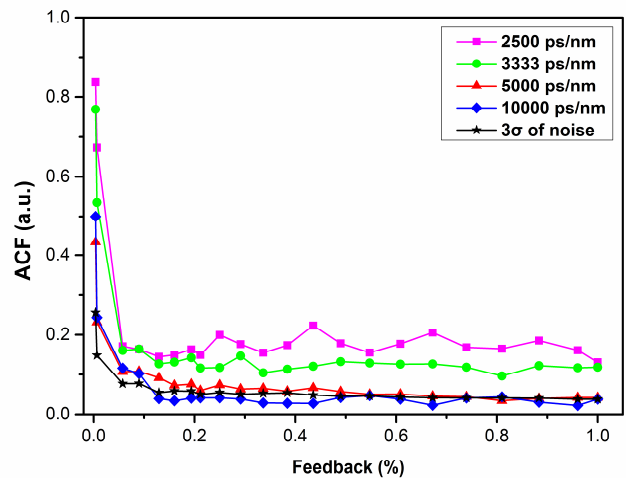

Fig. 6. (Color online) Numerical results of the TDS as function of feedback level obtained at different dispersion values of 2500ps/nm (squares), 3333ps/nm (dots), 5000ps/nm (triangles) and 10000ps/nm (diamonds). $I=1.5I_{th}$. $\sigma$ is the standard deviation of the background noise of ACF traces.

## III. EXPERIMENTAL RESULTS

The experimental arrangement of the CFBG feedback chaos is shown in Fig. 7. A distributed-feedback laser diode (DFB-LD) subject to dispersive feedback from a CFBG is shown in the dashed box. A polarization controller is used to match the polarization of the feedback light. A variable optical attenuator is utilized to adjust the feedback strength which is defined as the power ratio of the feedback light to the laser output. The DFB laser (Eblana, EP-1550-DM) is biased at 20mA (1.8 times threshold current) and has an output power of 1mW. The center wavelength is slightly tuned through a precise temperature controller (ILX Lightwave, LDT-5412). The CFBG has a length of 10 cm and a dispersion of approximately 2000ps/nm, which is measured with phase shift method [32]. The round-trip time between the laser and the grating is 43.3ns. For comparison, we also implement experiment of mirror feedback by replacing the grating with a fiber mirror. In this case, the feedback time delay is about 40.6ns which is close to that of CFBG feedback. In experiments, both the generated chaotic light and the feedback light are measured and analyzed. The time series and power spectrum are measured by a 36-GHz real-time oscilloscope (LeCroy, LabMaster10-36Zi) and a spectrum analyzer (Rohde and Schwarz, FSW) with a 12-GHz photodetector (Newport 1554-B). The optical spectrum of the laser is measured by an optical spectrum analyzer (APEX, AP2041-B) with a 5-MHz resolution. It is mentioned that, unlike the uniform FBG feedback [27], the CFBG response without filtering effects induces a low optical power loss and then the feedback system does not require optical amplification for feedback light, which simplifies the hardware configuration.

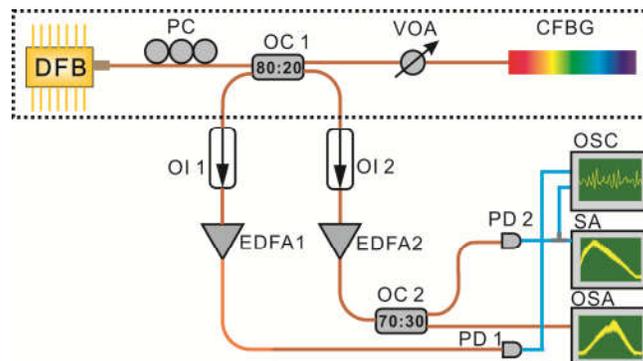

Fig. 7. Experimental setup of a DFB semiconductor laser with dispersive feedback from a CFBG. PC: polarization controller; OC: optical coupler; OI: optical isolator; VOA: variable optical attenuator; EDFA: Erbium-doped fiber amplifier; PD: photodetector; OSC: real-time oscilloscope; SA: spectrum analyzer; OSA: optical spectrum analyzer.

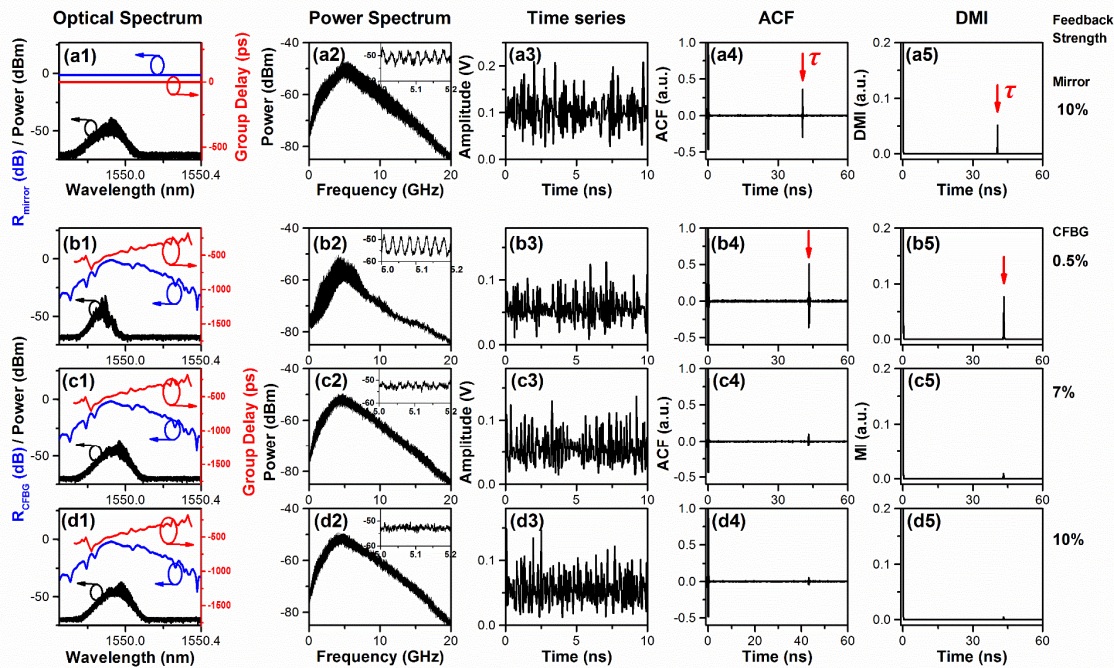

Fig. 8. Experimentally obtained optical spectra (column 1), power spectra (column 2), time series (column 3), ACF traces (column 4) and DMI traces (column 5): with mirror feedback at feedback strength of 10% (a1-a5), and with CFBG feedback separately at strength of 0.5% (b1-b5), 7% (c1-c5) and 10% (d1-d5). In the first column the reflection and delay spectrum of CFBG are plotted with optical spectrum together. In the second column, the insets plot the magnified local power spectrum in a range of 5-5.2GHz around the laser relaxation oscillation frequency.

Figure 8 demonstrates the evolution of TDS of CFBG feedback as feedback strength increases, in which columns 1-5 list optical spectra, power spectra, time series, ACF and DMI of the time series, respectively, from left to right. Note that the reflection spectrum and the group delay spectrum of CFBG are plotted in the first column with optical spectrum together. For comparison, we first show results of mirror feedback in the first row, which are obtained at intensity feedback level of 10%. As shown by the inset in Fig. 8(a2), the power spectrum exhibits a periodic fluctuation with a period of about 24MHz which is equal to $1/\tau$. This spectral periodic fluctuation, together with the ACF and DMI peaks in Figs. 8(a4) and 8(a5), clearly exhibits the TDS. The rows 2-4 in Fig.8 plot the results of CFBG feedback separately obtained at different intensity feedback levels of 0.5%, 7% and 10%. It is found from the ACF and DMI traces that the TDS gradually decreases as feedback level increases. Quantitatively, the ACF peak at dispersive feedback level of 10% is 0.04, which is decreased by 90% compared to the same level mirror feedback. The corresponding DMI peak is reduced by 92%. Depicted by the inset in Fig. 8(d2), the power spectrum no longer has the periodic modulation and thus the TDS in frequency domain is also eliminated.

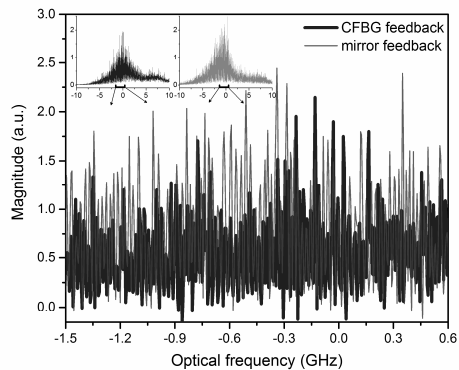

Fig. 9. Experimental local optical spectra of laser subject to mirror feedback (gray) and CFBG feedback (black) with intensity feedback level of 10%. The whole spectra are plotted in the insets which are corresponding to Figs. 8(a1) and 8(d1).

We take the case of feedback level of 10% as example to show the change of ECMs caused by the dispersive feedback. Fig. 9 magnifies a central frequency band of the optical spectra which are plotted in Figs. 8(a1) and 8(d1). As shown in the gray line, the mirror-feedback laser has many ECMs with a mode separation of about 24MHz in spite of that a few modes are absent due to mode hopping. By comparison, for CFBG-feedback laser as depicted by the black line, most spectral lines are suppressed. The remained spectral lines or modes have irregular mode spacing. These experimental results are in agreement with the numerical result in Fig. 3.

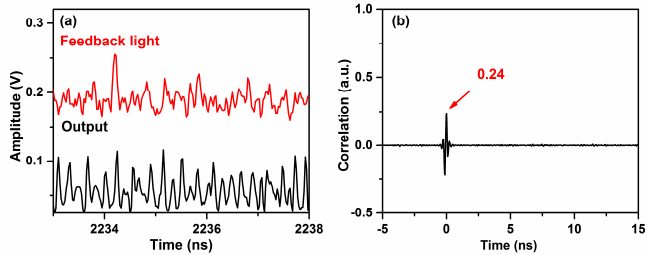

Fig. 10. (Color online) (a) Experimental time series of the laser output and the dispersive feedback light from CFBG and (b) their correlation trace, obtained with an intensity feedback level of 10%.

Further, we measure the temporal waveforms of the laser output and the feedback light from the CFBG to analyze their correlation, and plot the results in Fig.10. Note that, the laser output and the feedback light are the incident light and the reflected light of the CFBG, respectively. As shown in Fig. 10(a), compared to the waveform of laser output (black), the waveform of the CFBG feedback light (gray or red online) is broadened in temporal domain. This is no other than the result of dispersion. Fig. 10(b) plots the cross-correlation traces between the two temporal waveforms; the peak means a correlation coefficient of 0.24. This correlation coefficient is greatly smaller than that between incident and reflected light of mirror which is equal to 1. In addition, the correlation coefficient between the numerical laser output and feedback light corresponding to Fig. 2(c1) is calculated as 0.08. Therefore, the dispersive feedback is a kind of nonlinear feedback although it does not have filtering effects.

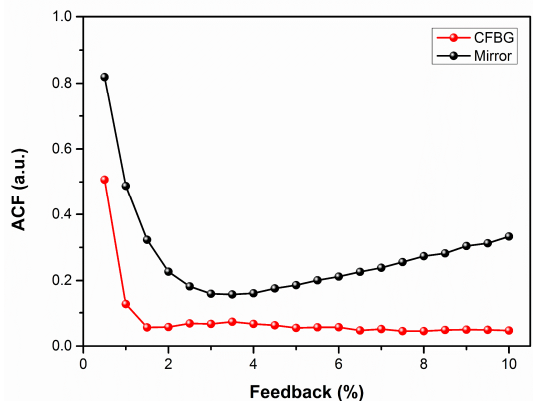

Fig. 11. (color online) Experimentally obtained ACF value at the delay time as function of feedback strength. Black: mirror feedback; gray (red): CFBG feedback. $I=1.8I_{th}$.

Figure 11 shows the experimentally measured TDS as function of feedback level. The black and gray (red online) lines represent the results of mirror feedback and CFBG feedback, respectively. Comparison shows that the TDS of dispersive feedback is greatly lower than that of mirror feedback at the same feedback strength. Moreover, the tendencies of TDS evolution are different. As feedback strength increases, the TDS of mirror feedback gradually decreases to a knee point at about 3% intensity feedback and then increases again. By comparison, the TDS of grating feedback decreases more rapidly to a knee point at about 1.5% intensity feedback, and then continues to decrease slowly down to a plateau as further increase of feedback strength. This experimental tendency agrees with the numerical results shown in Fig. 6. From the viewpoint of laser dynamics, for weak feedback on the left of the knee points, the laser mostly exhibits a periodic or quasi-periodic oscillation which mainly contributes to the autocorrelation peak at the delay time, i.e. TDS. As feedback level increases, the periodicity in dynamics is depressed and thus the TDS decreases. When the feedback is increased to the right of the knee point, the laser is chaotic and then the TDS mainly originates from the resonance of external cavity. For mirror feedback, increasing feedback level enhances the external cavity resonance and thus increases the TDS. For chirped grating feedback, the resonance is destroyed and then the TDS is depressed or eliminated and never regrow up as increasing feedback strength.

## IV. Conclusion

In summary, we numerically and experimentally investigate the time-delay signature of a semiconductor laser subject to dispersive optical feedback from a CFBG. Different from mirror, CFBG provides additional frequency-dependent delay induced by dispersion. As a result, the external-cavity modes have irregular mode separation rather than a fixed separation induced by mirror feedback. Numerical results show that the CFBG feedback leads to a similar quasi-period route to chaos with increases of feedback, compared with mirror feedback. But the TDS is greatly depressed and even eliminated. The elimination of TDS is confirmed by our experiments. By using a CFBG with dispersion of 2000ps/nm, the TDS is decreased by 90% to about 0.04 compared with mirror feedback. Moreover, the TDS evolution of the CFBG feedback is studied both in simulation and in experiments. The results show that the TDS evolution is quite different from mirror feedback: the TDS decreases more quickly down to a lower plateau (even ACF background noise level) and never grows again. This evolution tendency is also different from that of FBG feedback, of which the TDS first decreases to a minimal value and then increases again as feedback strength increases. The FBG feedback is actually a kind of filtered optical feedback, in which the laser is detuned to the edge of the main lobe of FBG spectrum and thus filtering effects is included. By comparison, the CFBG feedback has no filtering effects and is a paradigm of dispersive feedback. Additionally, the CFBG does not require optical amplification for feedback light, which simplifies the system configuration.